%% file: TOP-13-012_temp.tex
\pdfoutput=1

\documentclass[11pt,twoside,a4paper,cmspaper,final,collab]{cms-tdr}
\def\svnVersion{270479}\def\cmsCernNo{2013-037}\def\cmsCernDate{\today}\def\cmsMessage{Published in the Journal of High Energy Physics as \href{http://dx.doi.org/10.1007/JHEP11(2014)154}{\doi{10.1007/JHEP11(2014)154.}}}

\begin{document}\cmsNoteHeader{TOP-13-012}

\hyphenation{had-ron-i-za-tion}
\hyphenation{cal-or-i-me-ter}
\hyphenation{de-vices}

\RCS$Revision: 268727 $
\RCS$HeadURL: svn+ssh://svn.cern.ch/reps/tdr2/papers/TOP-13-012/trunk/TOP-13-012.tex $
\RCS$Id: TOP-13-012.tex 268727 2014-11-21 17:58:51Z keaveney $
\newcommand{\tttt}{\ensuremath{\ttbar\ttbar}\xspace} 
\newcommand{\ourLumi}{19.6\fbinv\xspace}
\newcommand{\ttbb}{\ttbar\bbbar}
\newcommand{\ttll}{\ttbar\it{LF}}
\newcommand{\ttjj}{\ensuremath{\ttbar jj}\xspace}
\newcommand{\sttbb}{\sigma(\ttbb)\xspace}
\newcommand{\sttjj}{\sigma(\ttjj)\xspace}

\newcommand{\subb}[1]{\ensuremath{_{\text{#1}}}}

\cmsNoteHeader{TOP-13-012} 
\title{Search for standard model production of four top quarks in the lepton + jets channel in pp collisions at $\sqrt{s}= 8$\TeV}

\date{\today}

\abstract{
A search is presented for standard model (SM) production of four top quarks (\tttt) in pp collisions in the lepton + jets channel. The data correspond to an integrated luminosity of 19.6\fbinv recorded at a centre-of-mass energy of 8\TeV with the CMS detector at the CERN LHC. The expected cross section for SM \tttt production is $\sigma^{\mathrm{SM}}_{\tttt} \approx 1\unit{fb}$. A combination of kinematic reconstruction and multivariate techniques is used to distinguish between the small signal and large background. The data are consistent with expectations of the SM, and an upper limit of 32\unit{fb} is set at a 95\% confidence level on the cross section for producing four top quarks in the SM, where a limit of $32\pm{17}$\unit{fb} is expected.}

\hypersetup{%
pdfauthor={CMS Collaboration},%
pdftitle={Search for standard model production of four top quarks in the lepton + jets channel in pp collisions at sqrt(s) = 8 TeV},%
pdfsubject={CMS},%
pdfkeywords={CMS, physics}}

\maketitle 

\section{Introduction}
\begin{figure}[h!]
\begin{center}
\subfloat{
\includegraphics[width=0.35\textwidth]{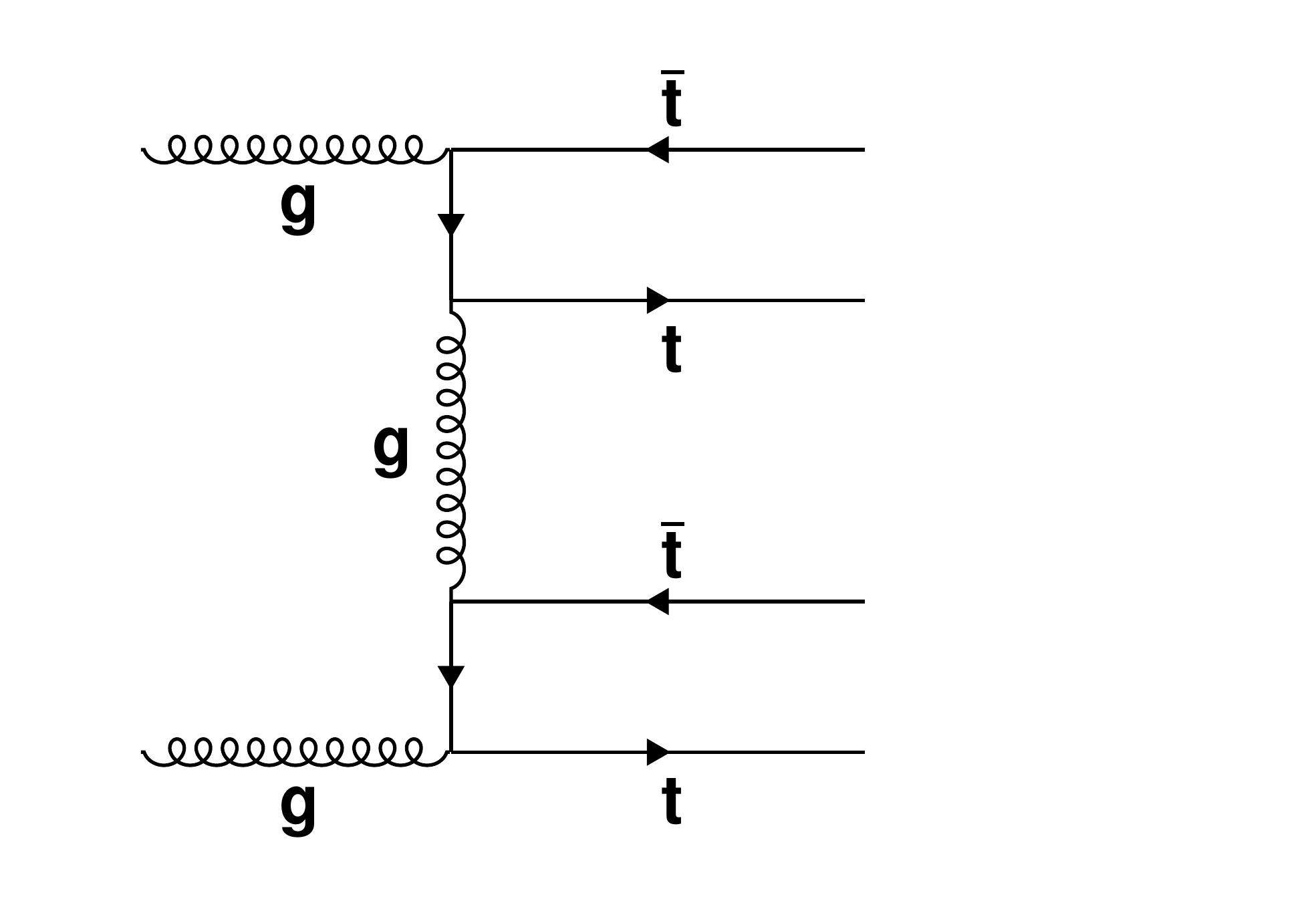}}
\subfloat{
\includegraphics[width=0.35\textwidth]{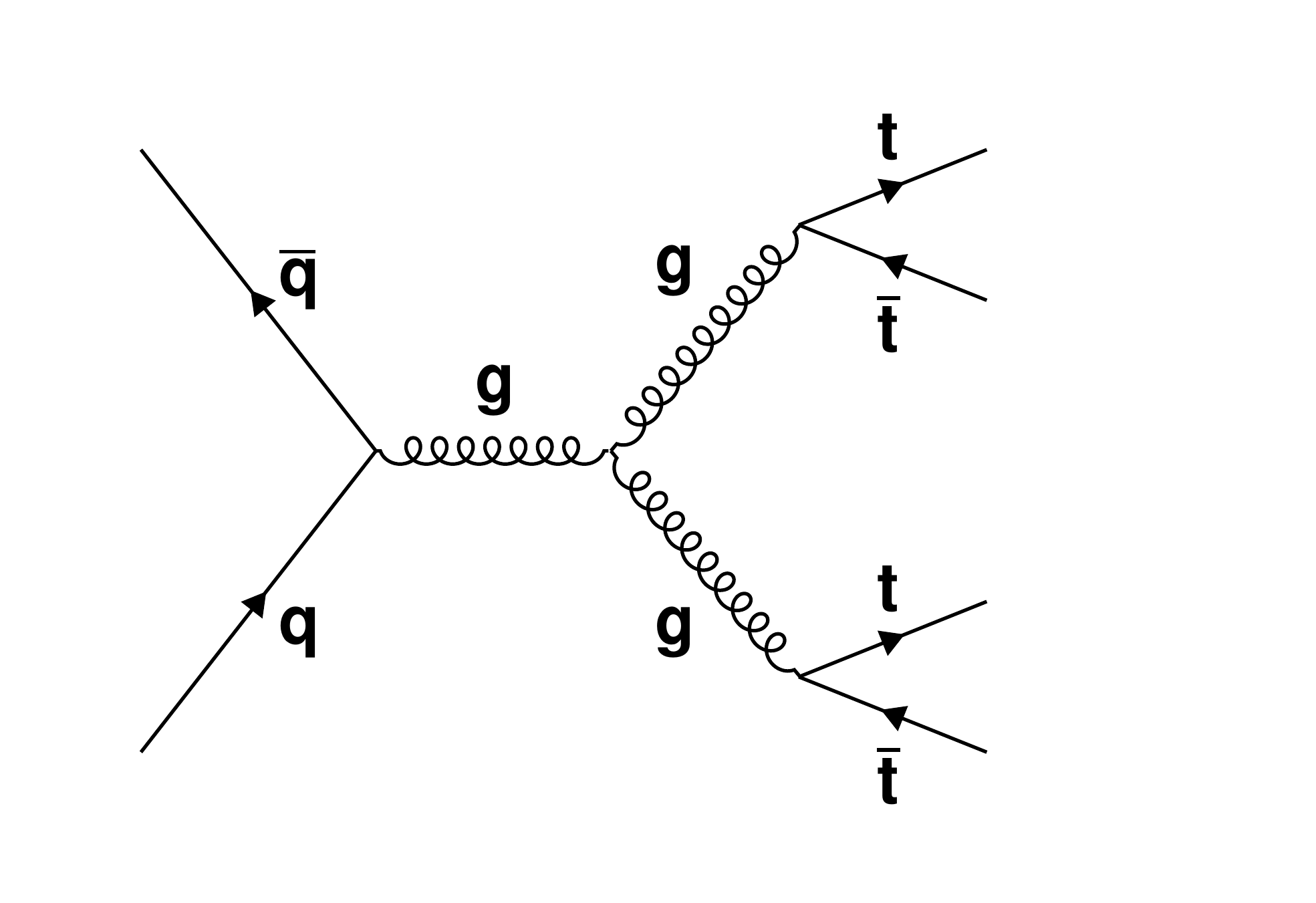}}
\end{center}
\caption{Leading-order Feynman diagrams for \tttt production in the SM from gluon-gluon fusion (left) and quark-antiquark annihilation (right).}
\label{fig:tttt_SM}
\end{figure}

Since its discovery in 1995 at the Fermilab Tevatron \cite{Abe:1995hr,Abachi:1995iq}, the top quark has been studied primarily using events containing top quark-antiquark pairs (\ttbar) and events containing a single top quark. With the larger centre-of-mass energy and luminosity of the CERN LHC, the study of more rare processes involving top quarks becomes possible. One such process is the production of four top quarks (\tttt). In the standard model (SM), \tttt production proceeds via gluon-gluon fusion or quark-antiquark annihilation. Feynman diagrams contributing to this process at leading order (LO) are shown in Fig.~\ref{fig:tttt_SM}. The cross section for SM \tttt production at the LHC is predicted, at LO, to be extremely small: $\sigma^{\text{SM}}_{\tttt} \approx$ 1\unit{fb} at $\sqrt{s} = 8 $ \TeV \cite{threeTop}. Next-to-leading-order (NLO) corrections increase the cross section by as much as 30\% \cite{ttttNLO}. The main background is due to \ttbar production, a process that has a cross section more than five orders of magnitude larger \cite{PhysRevLett.110.252004} and is one of the reasons that a \tttt signal has not yet been observed.

As the data used in this paper correspond to an integrated luminosity of \ourLumi, there are $\approx$20 \tttt events expected in the data. Because of the very large \ttbar background, the direct observation of events leading to a measurement of $\sigma_{\tttt}$ is unlikely.
However, in many models beyond the SM (BSM) involving massive coloured bosons, Higgs boson or top quark compositeness, or extra dimensions, $\sigma_{\tttt}$ is enhanced \cite{ttttNLO,AguilarSaavedra:2011ck,Plehn:1173428,GoncalvesNetto:2012nt,Fuks:2012im,Calvet:2012rk,Dev:2014yca}. In some supersymmetric extensions of the SM, \tttt final states can also be produced via cascade decays of coloured supersymmetric particles such as squarks and gluinos \cite{Toharia:2005gm}. In certain regions of BSM parameter space, these final states have kinematics similar to those of SM \tttt production. In such cases, reinterpretation of an upper limit on SM production of \tttt has the potential to constrain BSM theories. Moreover, in direct searches for these BSM signatures, SM production of \tttt can be a background. Hence, experimental constraints on $\sigma_{\tttt}$ have the potential to enhance the discovery reach of such searches.

This paper presents a search for SM production of \tttt in events that contain a single lepton ($\ell$) and multiple jets. Signal events are sought in final states with a single muon ($\mu$) or a single electron (e), in what are termed $\mu$ + jets and e + jets channels. The muon or electron originates either from the direct decay of a W boson or from the leptonic decay of a $\tau$ lepton in $ \cPqt \to \cPqb\PW$, $\PW \to \tau\nu_\tau$, $\tau \to \ell \nu_\ell \nu_\tau$. The chosen final state has a larger branching ratio ($\approx$41\%) than the zero-lepton ($\approx$30\%), two-lepton ($\approx$22\%) or three- and four-lepton ($\approx$6\%) final states, when only muons and electrons are considered as final-state leptons. Kinematic reconstruction techniques and multivariate analyses (MVA) are used to discriminate the \tttt signal from the \ttbar background.

\section{The CMS detector}
\label{sec:CMS}
The central feature of the CMS apparatus is a superconducting solenoid of 6\unit{m} internal diameter, providing a magnetic field of 3.8\unit{T}. The field volume contains a silicon pixel and strip tracker, an electromagnetic calorimeter (ECAL), and a hadron calorimeter (HCAL). Charged particles are reconstructed in the pseudorapidity range $\abs{\eta} < 2.5$. The tracker has a track-finding efficiency of more than 99\% for muons with transverse momenta \pt more than 1\GeV. The ECAL is a fine-grained hermetic calorimeter, consisting of lead tungstate crystals arranged in a quasi-projective geometry, and distributed in the barrel region of $\abs{\eta}< 1.48$ and in two endcaps that extend up to $\abs{\eta}< 3.0$. The HCAL barrel and endcaps are sampling calorimeters that consist of brass and scintillator plates, covering the region of $\abs{\eta} < 3.0$. In addition to barrel and endcap detectors, CMS has extensive forward calorimetry. Muons are measured in gas-ionisation detectors embedded in the steel flux-return yoke outside the solenoid. A more detailed description of the CMS detector, together with a definition of the coordinate system used and the relevant kinematic variables, can be found in Ref.~\cite{Chatrchyan:2008zzk}.

\section{Data and simulation\label{sec:datasim}}
The data are collected using triggers based on the presence of a muon candidate with $\pt >24$\GeV, or an electron candidate with $\pt > 27\GeV$.
The signal process is modelled at LO using the \MADGRAPH (v5.1.3.30) Monte Carlo (MC) generator~\cite{Alwall:2011uj}. The \ttbar background is modelled at LO with up to three additional partons in the zero-lepton, one-lepton and two-lepton final states, also using \MADGRAPH, as is the production of \ttbar + electroweak (EW) bosons and the production of EW bosons with additional partons. Single top quarks and any additional partons produced in the $t$- and $s$-channels, and in association with a W boson, are modelled at NLO using \POWHEG (v1.0 r1380)~\cite{Alioli:2009je,Alioli:2009jeErratum,Nason:2004rx,Frixione:2007vw,Alioli:2010xd}. The \ttbar + Higgs boson and diboson (WW, WZ and ZZ) processes are modelled at LO using \PYTHIA 6.426~\cite{Sjostrand:2006za}. The decays of $\tau$ leptons are simulated with \TAUOLA (v27.121.5)~\cite{Davidson:2010rw}. The simulation of additional initial-state and final-state radiation, and the fragmentation and hadronisation of quarks and gluons are performed using \PYTHIA  with the Z2* tune~\cite{Z2Star}. The (LO) CTEQ6L1~\cite{Pumplin:2002vw} set of parton distribution functions (PDF) is used with the \MADGRAPH  and \PYTHIA samples while the (NLO) CTEQ6M set is used with the \POWHEG samples. The generated events are passed through a full simulation of the CMS detector, based on the \GEANTfour package~\cite{Agostinelli2003250}. Next-to-leading-order or, when available, next-to-next-to-leading-order (NNLO) cross sections are used to normalise the predictions.

For all samples of simulated events, multiple minimum bias events generated with \PYTHIA are added to simulate the presence of additional proton-proton interactions in a single or neighbouring proton bunches (pileup). To refine the simulation, the simulated events are weighted to reproduce the distribution in the number of reconstructed vertices observed in data.

The cross section ratio $\sigma_{\ttbar\bbbar}/\sigma_{\ttbar jj}$ is measured by CMS to be $2.2 \pm 0.3\stat\pm0.5\syst$\%, where the $\ttbar$$jj$ process is defined as the production of a \ttbar pair with any two additional jets ($j$) \cite{CMS-PAS-TOP-13-010}. As the \MADGRAPH \ttbar simulation used in this analysis predicts $\sigma_{\ttbar\bbbar}/\sigma_{\ttbar jj}$ to be 1.2\%, the simulation is corrected to the ratio of cross sections measured in Ref. \cite{CMS-PAS-TOP-13-010}. To illustrate the composition of this dominant background, the sample is divided into two subcategories. The categories correspond to the production of \ttbar with additional light quarks or gluons ``\ttbar + ll/gg" and to the production of \ttbar with additional charm or bottom quarks ``\ttbar + \ccbar/\bbbar". The samples of electroweak bosons with additional partons and of single top quarks are grouped into a category termed ``EW", while samples of \ttbar + electroweak bosons, \ttbar decaying into final states containing zero or two leptons are grouped into a category termed ``\ttbar other".
\section{Event reconstruction and selections \label{sec:eventselection}}
Events are reconstructed using a particle-flow (PF) algorithm \cite{CMS-PAS-PFT-09-001,CMS:2010byl}. This proceeds by reconstructing and identifying each final-state particle using an optimised combination of all subdetector information. Each event is required to have at least one reconstructed vertex. The primary vertex is chosen as the vertex with the largest value of $\Sigma \pt^{2}$ of the tracks associated with that vertex. Additional selection criteria are applied to each event to reject events with features consistent with arising from detector noise and beam-gas interactions.

The energy of electrons is determined from a combination of the track momentum at the primary vertex, the corresponding ECAL energy cluster, and the energy sum of the reconstructed bremsstrahlung photons associated with the track. The energy of muons is obtained from the corresponding track momentum obtained in a combined fit to information from the inner silicon trackers and outer muon detectors. The energy of charged hadrons is determined from a combination of the track momentum and the corresponding ECAL and HCAL energies, corrected for the suppression of small signals, and calibrated for the non-linear response of the calorimeters. Finally, the energy of neutral hadrons is obtained from the corresponding calibrated ECAL and HCAL energies. As charged leptons originating from top quark decays are typically isolated from other particles, a variable ($I$\subb{rel}) is constructed to select lepton candidates based on their isolation. It is defined as the scalar sum of the \pt values of the particles reconstructed within an angle $\Delta R$ of the axis of the momentum of the lepton candidate, excluding the lepton candidate, divided by the \pt of the lepton candidate. The $\Delta R$ is defined as $\Delta R = \sqrt{(\Delta \eta)^{2} + (\Delta \phi)^{2}}$, where $\Delta \eta$ and $\Delta \phi$ are the differences in pseudorapidity and azimuthal angles between the lepton candidate and any other track or energy deposition. A muon or an electron candidate is rejected if $I$\subb{rel} is greater than or equal to 0.12 or 0.1, for respective values of $\Delta R$ of 0.4 and 0.3.

Jets are clustered from the reconstructed particles using the infrared- and collinear-safe anti-\kt algorithm \cite{Cacciari:2008gp}, with distance parameter $R=0.5$, as implemented in the \textsc{FastJet} package \cite{Cacciari:2011ma}. The jet momentum is defined by the vectorial sum of the momenta of all of the particles in each jet, and is found in the simulation to be within 5\% to 10\% of the true jet momentum, for the entire \pt spectrum of interest and detector acceptance \cite{Chatrchyan:2011ds}. Corrections to the jet energy scale (JES) and the jet energy resolution (JER) are obtained from the simulation and through in situ measurements of the energy balance of exclusive dijet and photon + jet events. An offset correction is applied to take into account the extra energy clustered into the jets from pileup. Muons, electrons, and charged hadrons originating from pileup interactions are not included in the jet reconstruction. Missing transverse energy (\MET) is defined as the magnitude of the vectorial sum of the \PT of all the selected jets and leptons in the event. Charged hadrons originating from pileup interactions are also not included in the reconstruction of \MET. Jets are classified as b quark jets through their probability of originating from the hadronisation of bottom quarks, using the combined secondary vertex (CSV) b tagging algorithm, which combines information from the significance of the track impact parameter, the jet kinematics, and the presence of a secondary vertex within the jet \cite{CMS-PAS-BTV-13-001}.

To preferentially select \tttt events while suppressing backgrounds, events that pass the muon or electron trigger are required to pass baseline selections corresponding to the $\mu$ + jets or e + jets channels. The selections comprise a series of criteria applied to the objects in the offline-reconstructed event. The selections require the presence of one well-identified and isolated muon or electron \cite{Chatrchyan:2012xi,CMS-DP-2013-003} with $\pt >  30$\GeV and with respective muon or electron $\abs{\eta}< 2.1$ or 2.5. Jets are required to have $\pt>30\GeV$ and $\abs{\eta}<2.5$. All events are required to have the number of selected jets (N\subb{jets}) to be at least six. For a jet to be b tagged, it must pass a requirement of the CSV algorithm \cite{CMS-PAS-BTV-13-001} that provides a misidentification rate of $\approx$1\% for light quark and gluon jets, and corresponds to b tagging efficiencies of 40--75\%, depending on jet \pt and $\eta$.
Events are required to have the number of b-tagged jets ($N$\subb{btags}) to be at least two. The requirements on $N$\subb{jets} and $N$\subb{tags} strongly suppress background events arising from vector boson + jets and single top quark production. The \HT of an event is defined as the scalar sum of the \pt of all the selected jets in the event. Events are also required to have $\HT > 400\GeV$ and $\MET > 30\GeV$, which removes the residual background arising from multijet processes. Small corrections are applied to the simulated events to account for the differences between efficiencies in data and simulation for the above lepton and b tagging requirements.

\section{Event classification with an MVA algorithm\label{sec:bdt}}
To obtain the greatest possible discrimination between \tttt events and the dominant \ttbar background, variables sensitive to the different processes are exploited in a multivariate (MVA) discriminant. The selected variables are grouped into three categories based on the underlying physical characteristics that they exploit: content of top quarks, jet activity, and b quark jet content.

\subsection{Multiplicity of top quarks}
\label{sec:topcon}
The presence of multiple jet-decaying top quarks in \tttt events can be exploited to distinguish such events from \ttbar background, which contains only a single jet-decaying top quark. The challenge in the kinematic reconstruction of such top quarks in an event containing many jets is to find correct selections of three jets that arise from any single top quark when many incorrect three-jet combinations are possible. Such correctly selected combinations are referred to as ``correct trijets", while combinations containing one or more jets not originating from the same top quark are referred to as ``incorrect trijets".
The large number of incorrect trijets in signal and background events motivates the use of MVA methods as in Ref. \cite{Chatrchyan:2012fp} to distinguish between correct and incorrect trijets by combining information from a set of input variables.

Within the trijet, the dijet system originating from W boson decay is attributed to the two jets with smallest $\Delta R$ separation. The invariant mass of this dijet and the invariant mass of the trijet are used as input variables in the MVA. For correct trijets, these variables have respective values close to that of the W boson and of the top quark. The azimuthal separations between the trijet and dijet systems and between the trijet and the jet not selected for the dijet system are also used. These variables typically have smaller values for correct trijets. The ratio of the magnitude of the vectorial \pt sum to the scalar \pt sum of the trijet system has typically larger values for correct trijets and is included as an input variable. Finally, the discriminant of the CSV b tagging algorithm for the jet not selected in the dijet provides, for correct trijets, values expected for b quark jets and is therefore included. These variables are combined in a boosted decision tree algorithm (BDT\subb{trijet}) using the {TMVA} package \cite{Hocker:2007ht}. Of the simulated \ttbar events that pass the baseline selections, approximately 61\% have the trijet with the largest BDT\subb{trijet} discriminant value for the $\cPqt\to \text{trijet decay}$.

\begin{figure}[!htb]
\begin{center}
\subfigure[]{\includegraphics[width=.49\textwidth]{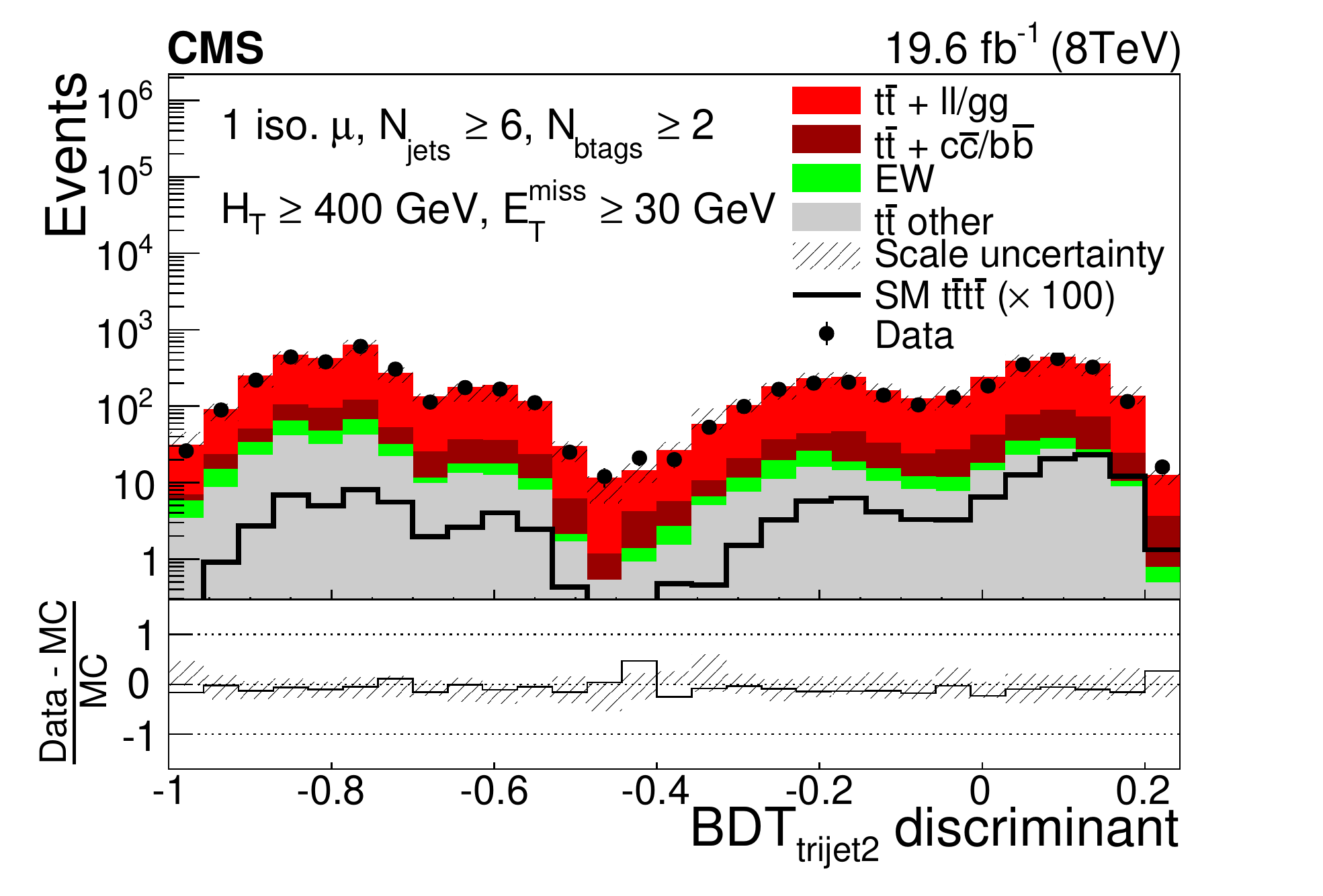}}
\subfigure[]{\includegraphics[width=.49\textwidth]{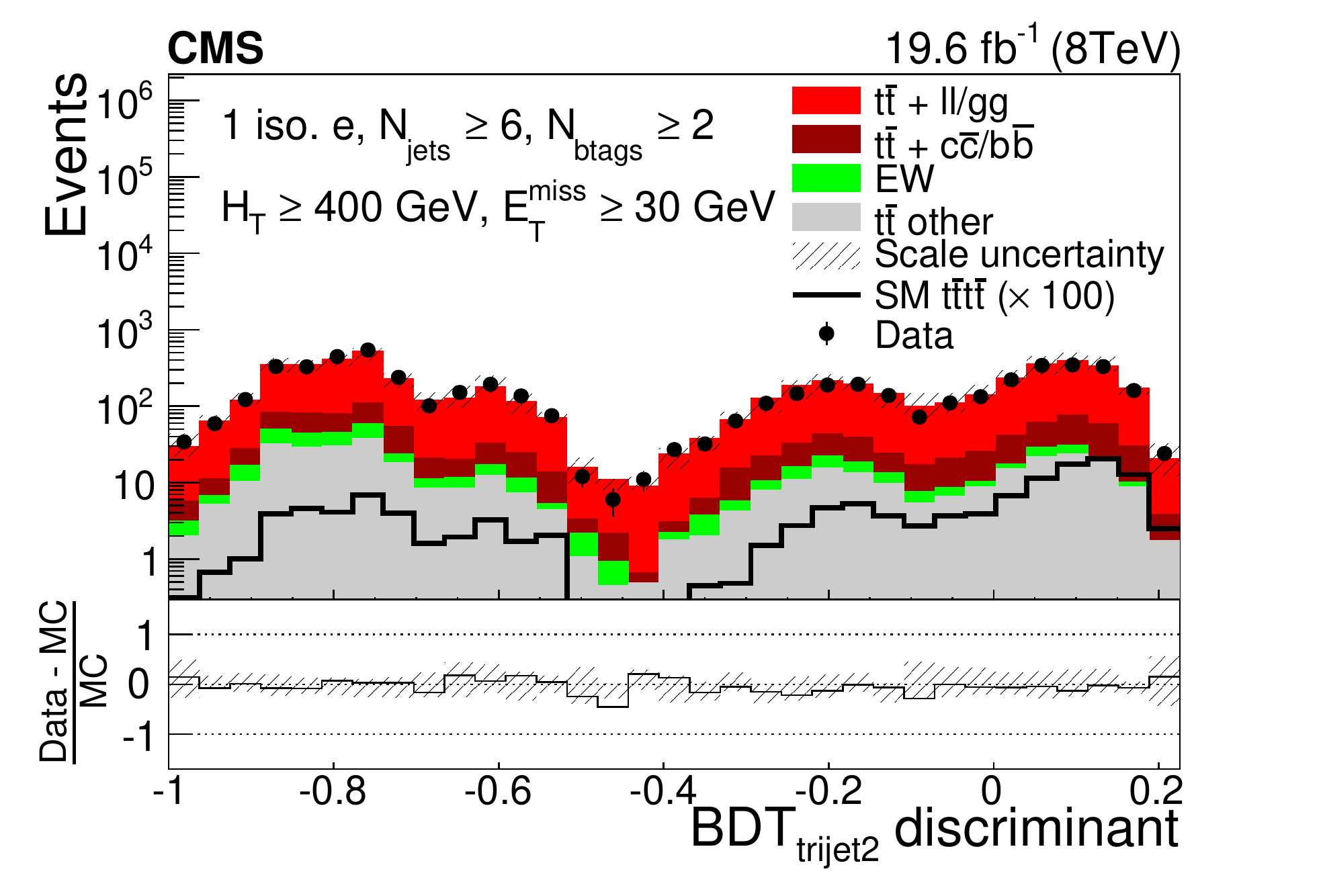}} \\
\subfigure[]{\includegraphics[width=.49\textwidth]{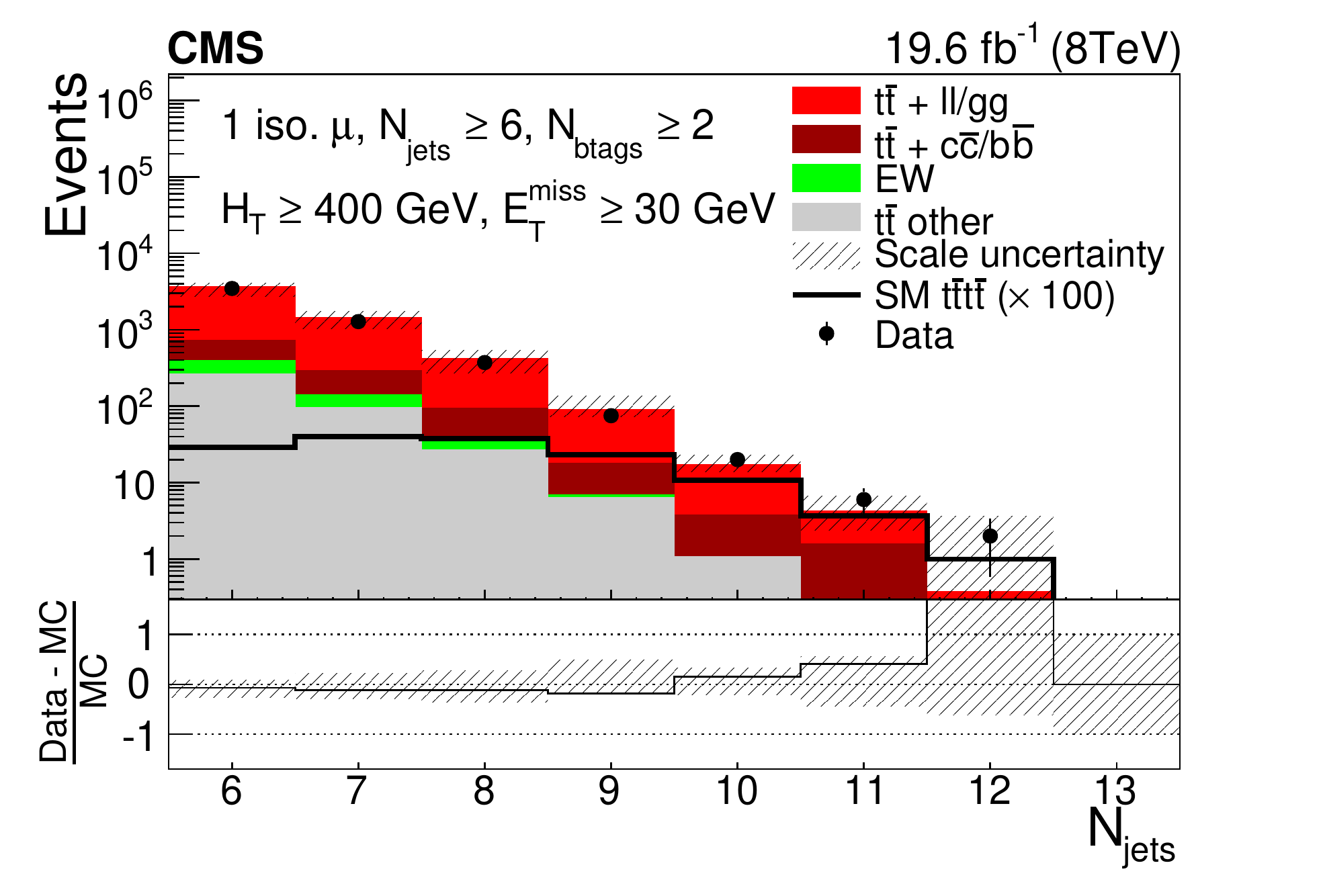}}
\subfigure[]{\includegraphics[width=.49\textwidth]{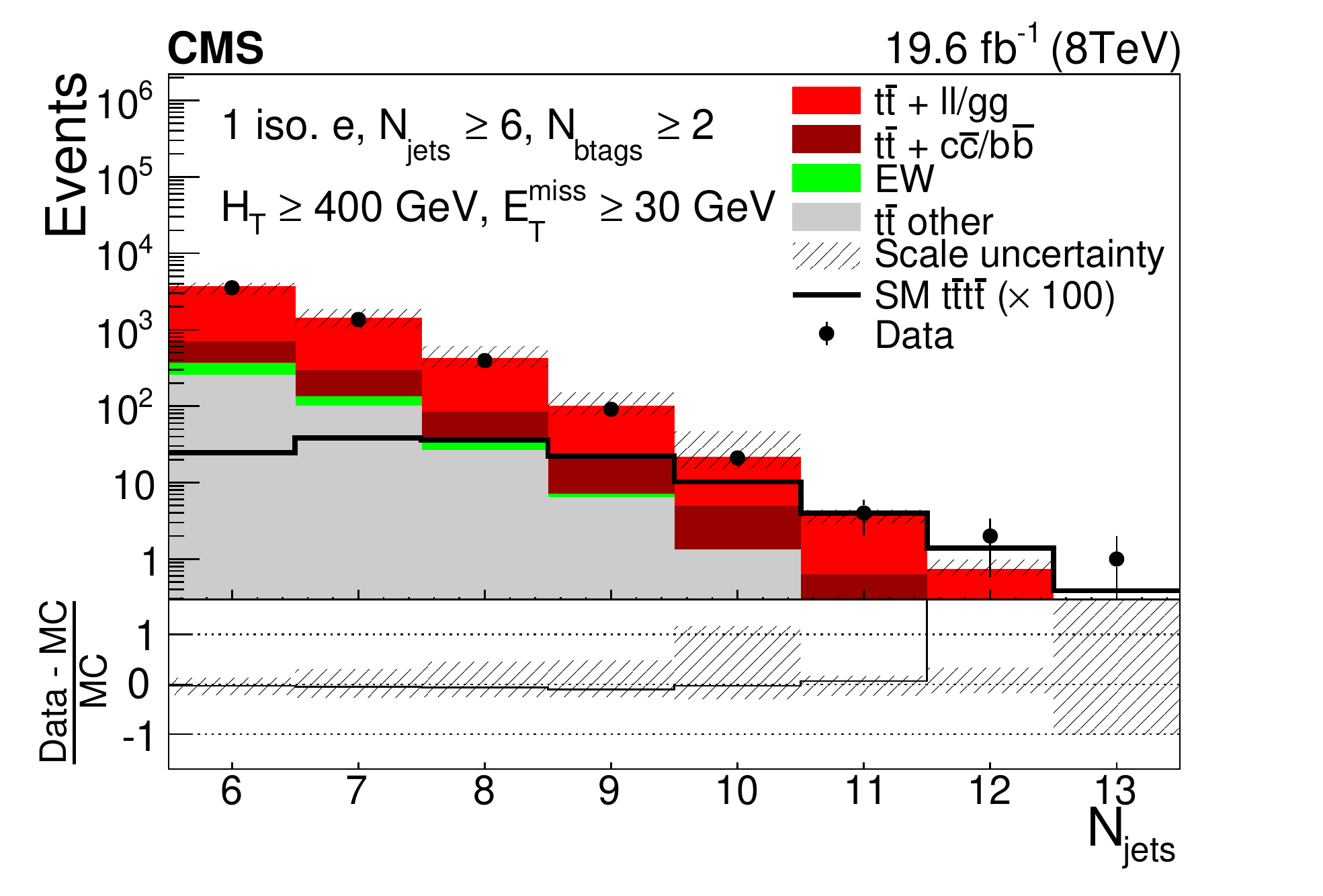}} \\
\subfigure[]{\includegraphics[width=.49\textwidth]{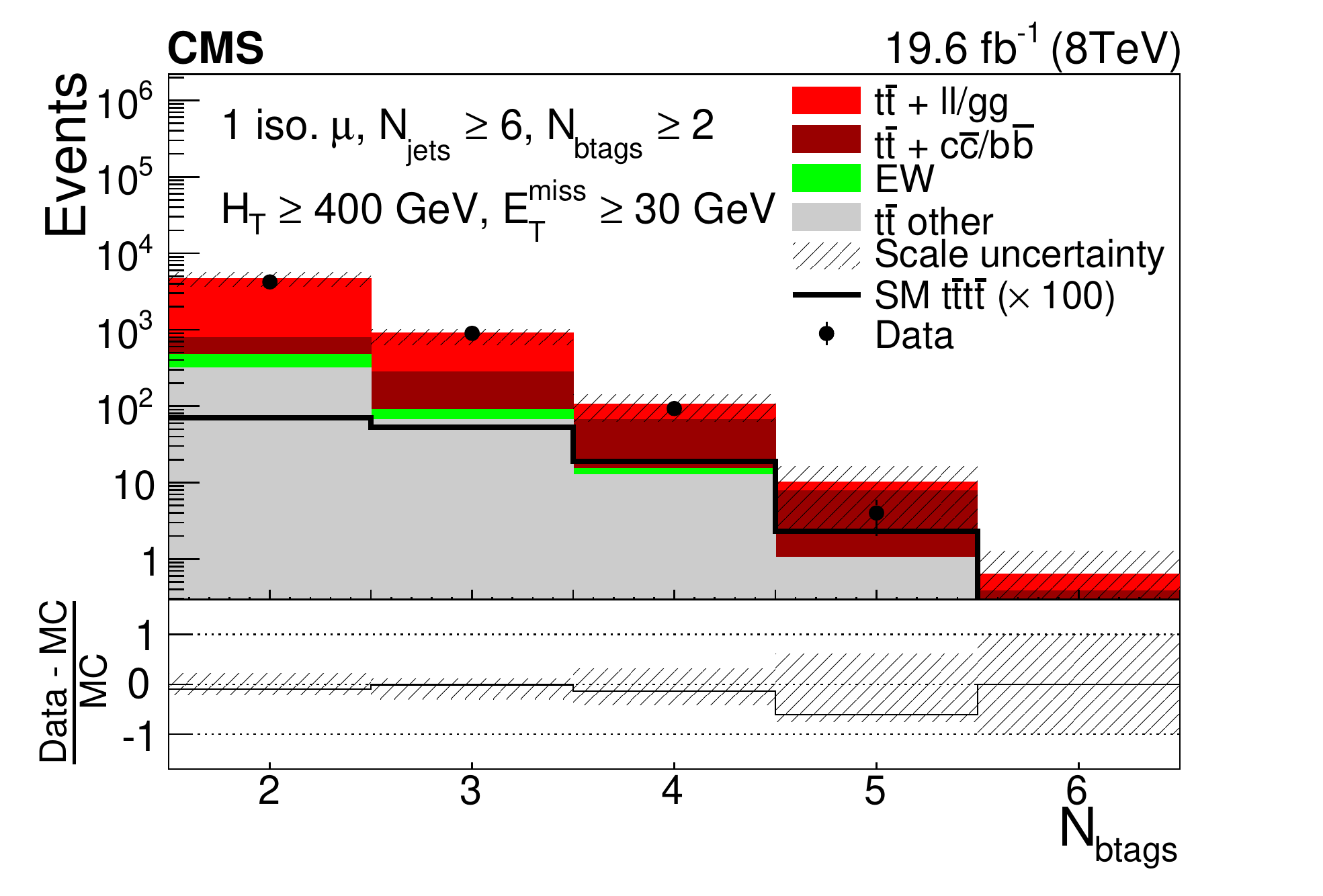}}
\subfigure[]{\includegraphics[width=.49\textwidth]{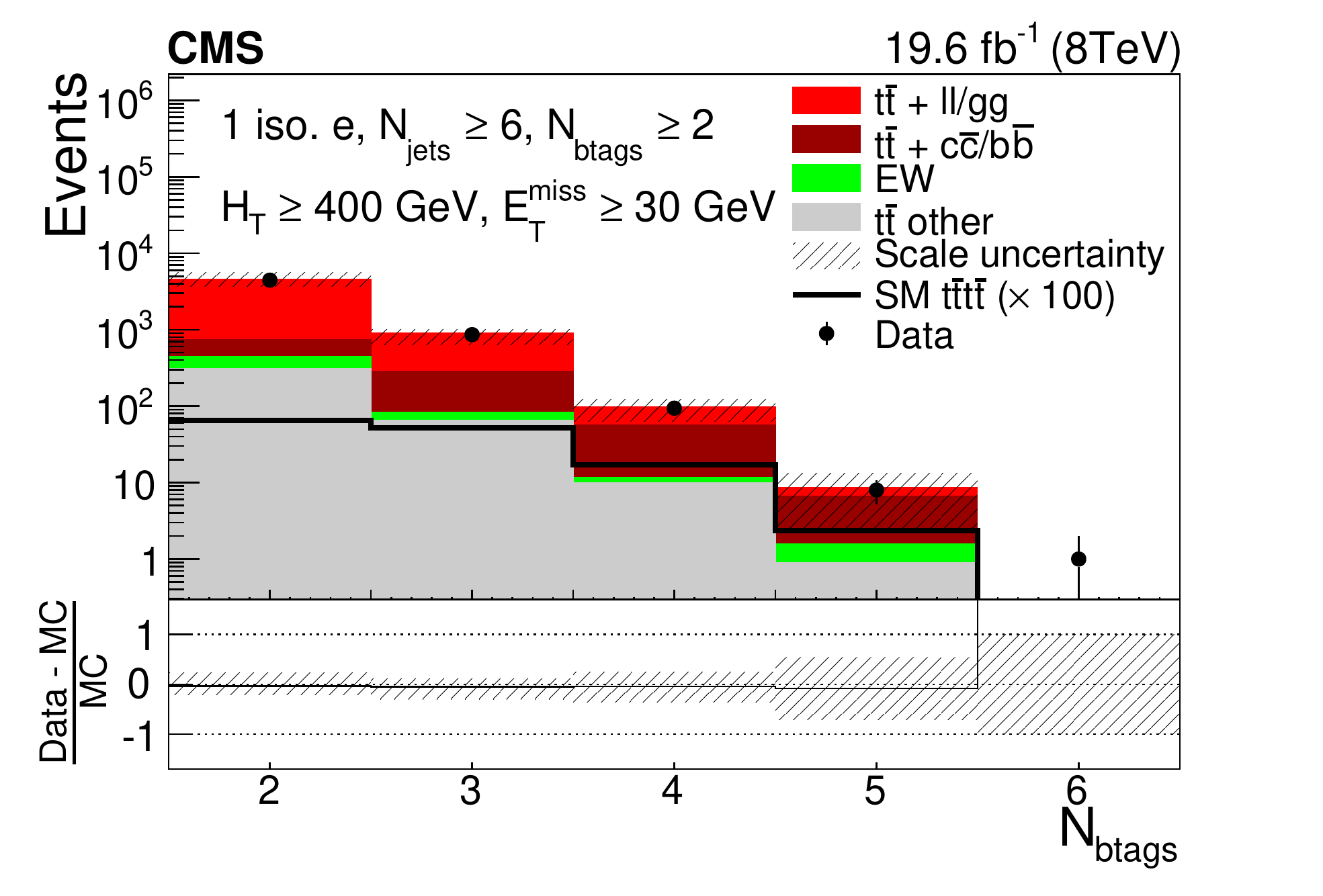}} \\
\caption{The distribution in the BDT\subb{trijet2} discriminant for the $\mu$ + jets and e + jets channels in (a) and (b), respectively, and the same for $N$\subb{jets} in (c) and (d), and for $N$\subb{btags} in (e) and (f). The ratios plotted at the bottom of each panel reflect the percent differences between data and MC events. The hatched areas show the changes in the calculated predictions produced by factors of two and one half changes in the factorisation and renormalisation scales in the \ttbar simulation.}
\label{fig:BC0}
\end{center}
\end{figure}

Following the baseline selection, it should be possible to reconstruct multiple all-jet decays of top quarks in \tttt events, but not in \ttbar events. The BDT\subb{trijet} distribution in the highest-scoring trijet discriminant has a similar form for \tttt and \ttbar events. However, the second-highest ranking trijet frequently reflects correct trijets in \tttt events and incorrect trijets in \ttbar events. Hence, the trijet with the largest value of BDT\subb{trijet} discriminant is removed from the event, and the BDT\subb{trijet} discriminant of the trijet of the remaining jets with highest value (BDT\subb{trijet2}) is used to distinguish between \tttt and \ttbar events. Distributions in this variable, in $N$\subb{jets} and in $N$\subb{btags} in data and simulation are shown in Fig. \ref{fig:BC0}. The uncertainty in ``scale" in the legend refers to the changes produced through changes of factors of two and one half in the factorisation and renormalisation scales of the calculation, as discussed in Section 6. The successful reconstruction of a third all-jet decay of a top quark is unlikely, and is only possible in the small fraction of events containing at least nine jets. Hence it provides negligible additional discriminating power and is not used.

The reduced event (RE) is constructed by subtracting the jets contained in the highest BDT\subb{trijet}-ranking trijet. In \ttbar events, the RE will typically contain only jets arising from $\cPqt\to\cPqb\ell\nu_{\ell}$ decays of the top quark, initial- and final-state radiation, the underlying event, and pileup interactions. Conversely, a \tttt RE can contain up to two all-jet top quark decays, and as a result numerous energetic jets. Two variables based on the RE are (i) $\HT^\mathrm{RE}$, \ie, the \HT of the RE and (ii) $M^{\mathrm{RE}}$, \ie, the invariant mass of the system comprising all the jets in the RE.

\subsection{Jet activity}
\label{sec:jetact}
Because \tttt events can contain up to ten hard jets from top quark decays, while \ttbar events contain up to four, the following variables based on jet activity of the event possess discrimination power: (i) $N$\subb{jets}, (ii) $\HT^\cPqb$, (iii) $\HT/H_\mathrm{p}$, (iv) $\HT^\text{ratio}$, (v) $p_{\mathrm{T5}}$, and (vi) $p_{\mathrm{T6}}$. The $\HT^\cPqb$ variable is defined to be the \HT of the b-tagged jets. In the $\HT/H\subb{p}$ ratio, $H\subb{p}$ is the scalar sum of the total momenta of the selected jets. The ratio of the \HT of the four leading jets to the \HT of the other jets is defined as $\HT^{\text{ratio}}$. The $p_{\mathrm{T5}}$ and $p_{\mathrm{T6}}$ variables represent, respectively, the \pt values of jets of 5th and 6th largest \pt. All these variables are used in the discriminant described in Section \ref{sec:evBDT}.

\subsection{Multiplicity of bottom quarks}
\label{sec:bcon}
The analysis assumes that the top quark decays with the SM branching ratio of $\mathcal{B}({\cPqt \to \cPqb\PW})=1$. Hence, \tttt events contain four bottom quarks from top quark decays whereas \ttbar events contain only two bottom quarks from top quark decays. Therefore the multiplicity of b-tagged jets is a potential source of discriminating power, and is also used in the discriminant discussed in Section \ref{sec:evBDT}.

\subsection{Event-level BDT}
\label{sec:evBDT}
\begin{figure}[!htb]
\begin{center}
\subfigure[]{\includegraphics[width=.49\textwidth]{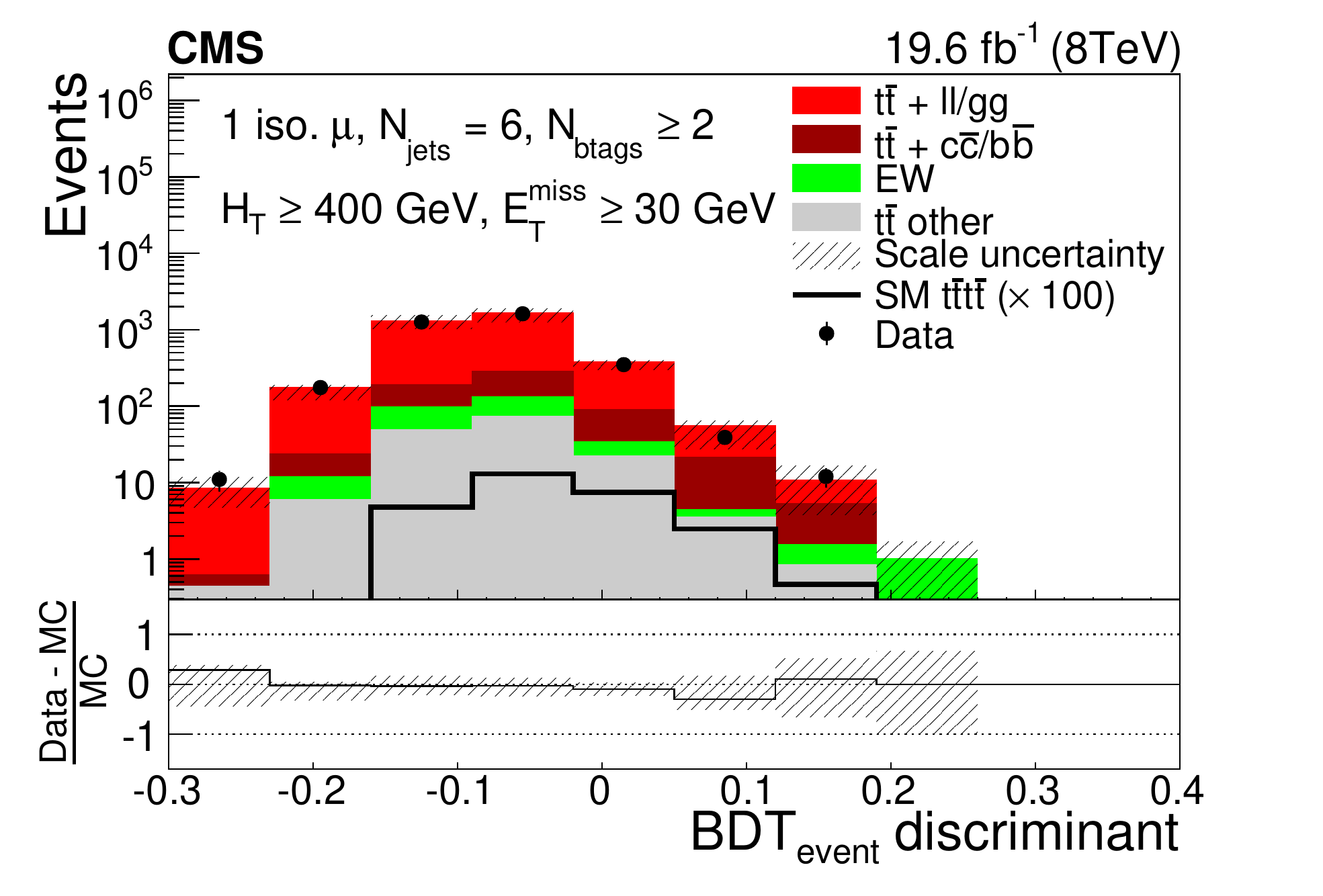}}
\subfigure[]{\includegraphics[width=.49\textwidth]{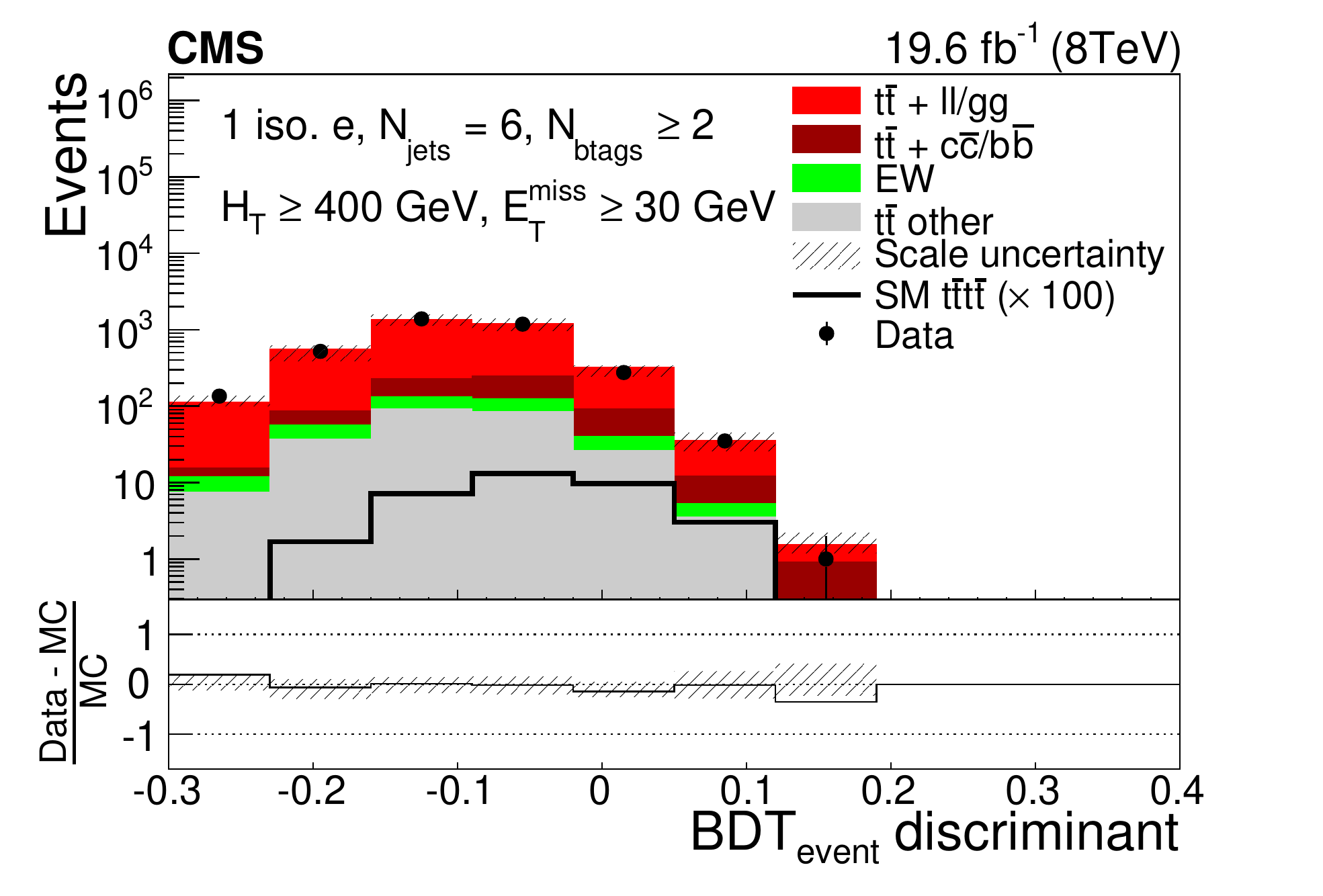}} \\
\subfigure[]{\includegraphics[width=.49\textwidth]{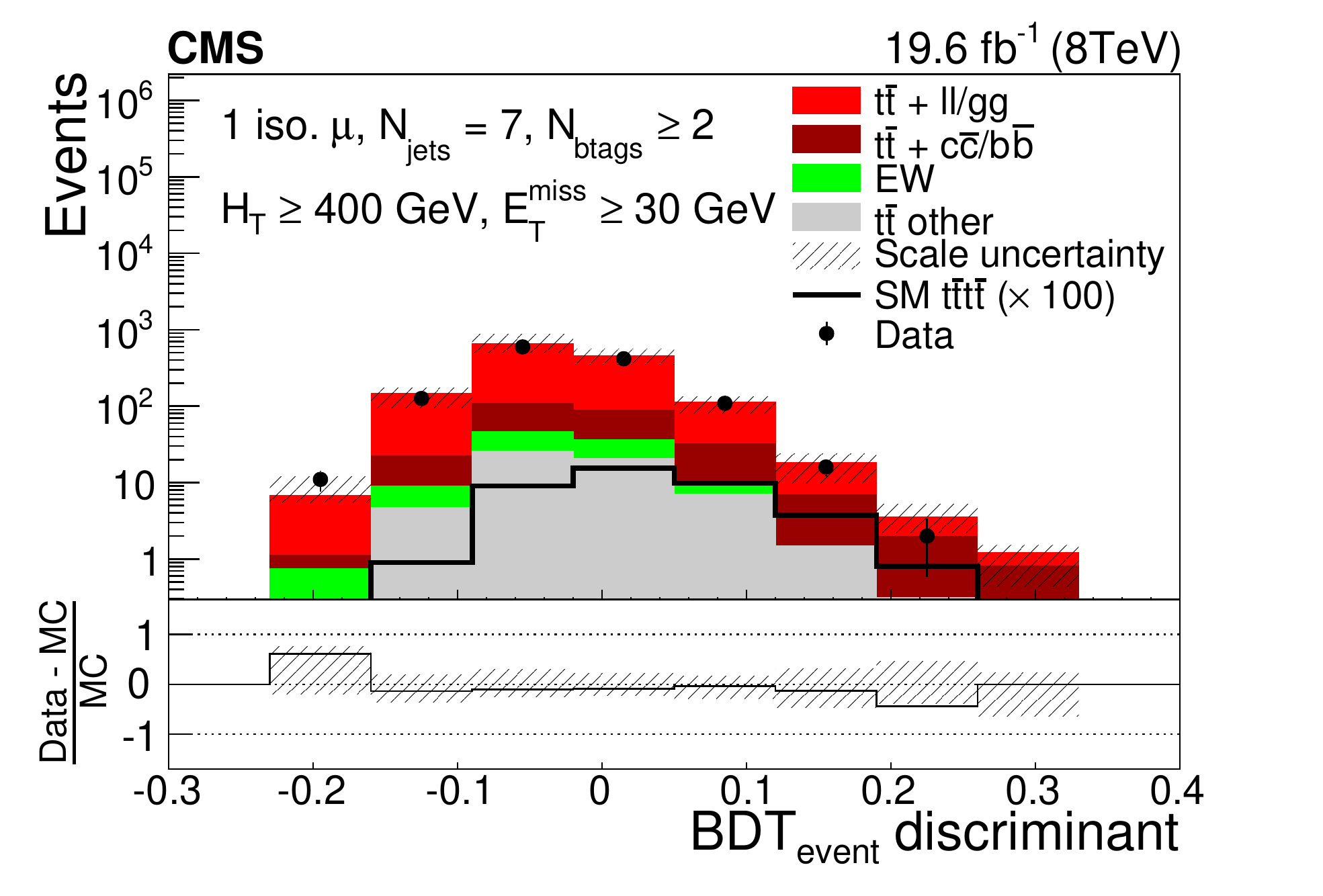}}
\subfigure[]{\includegraphics[width=.49\textwidth]{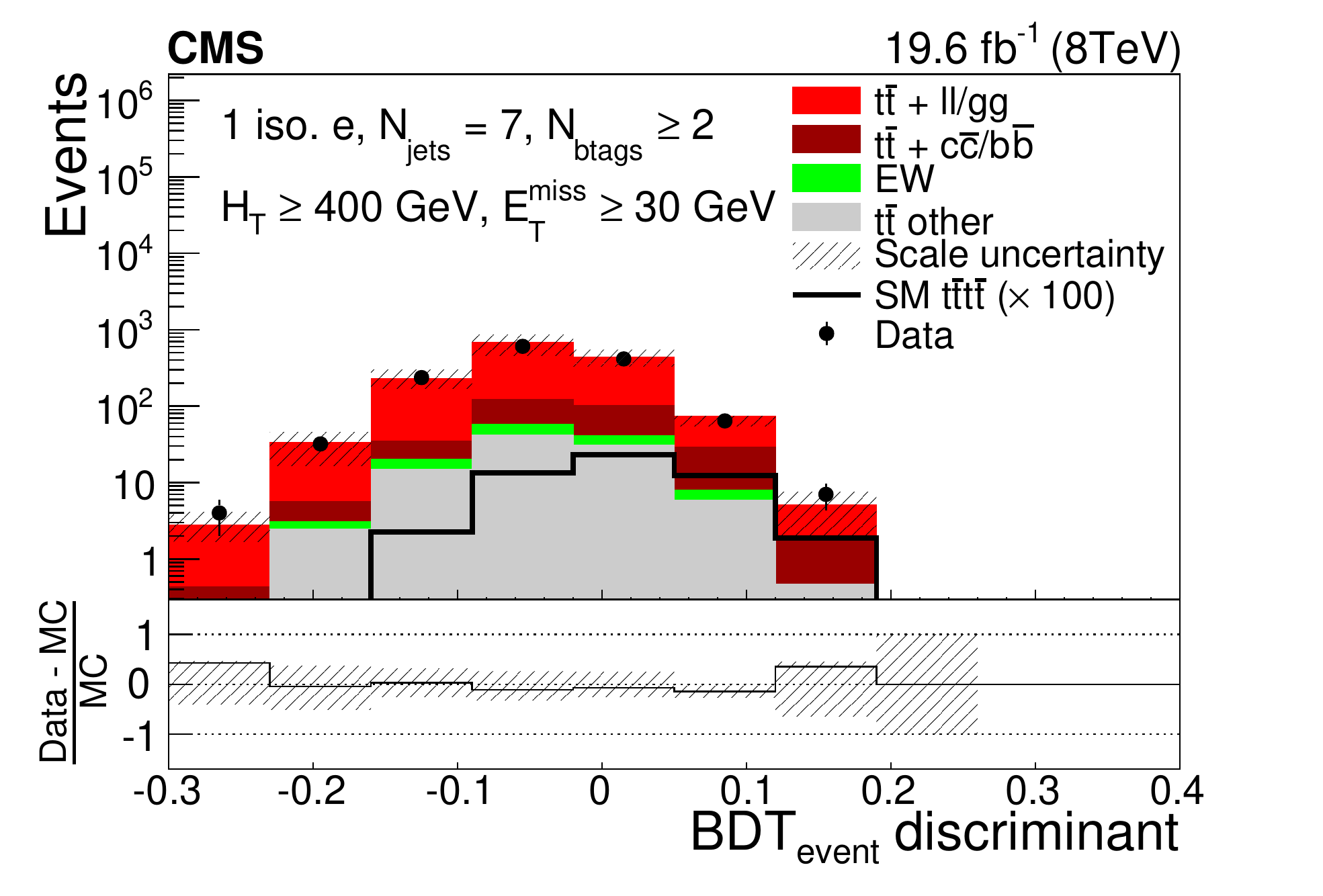}} \\
\subfigure[]{\includegraphics[width=.49\textwidth]{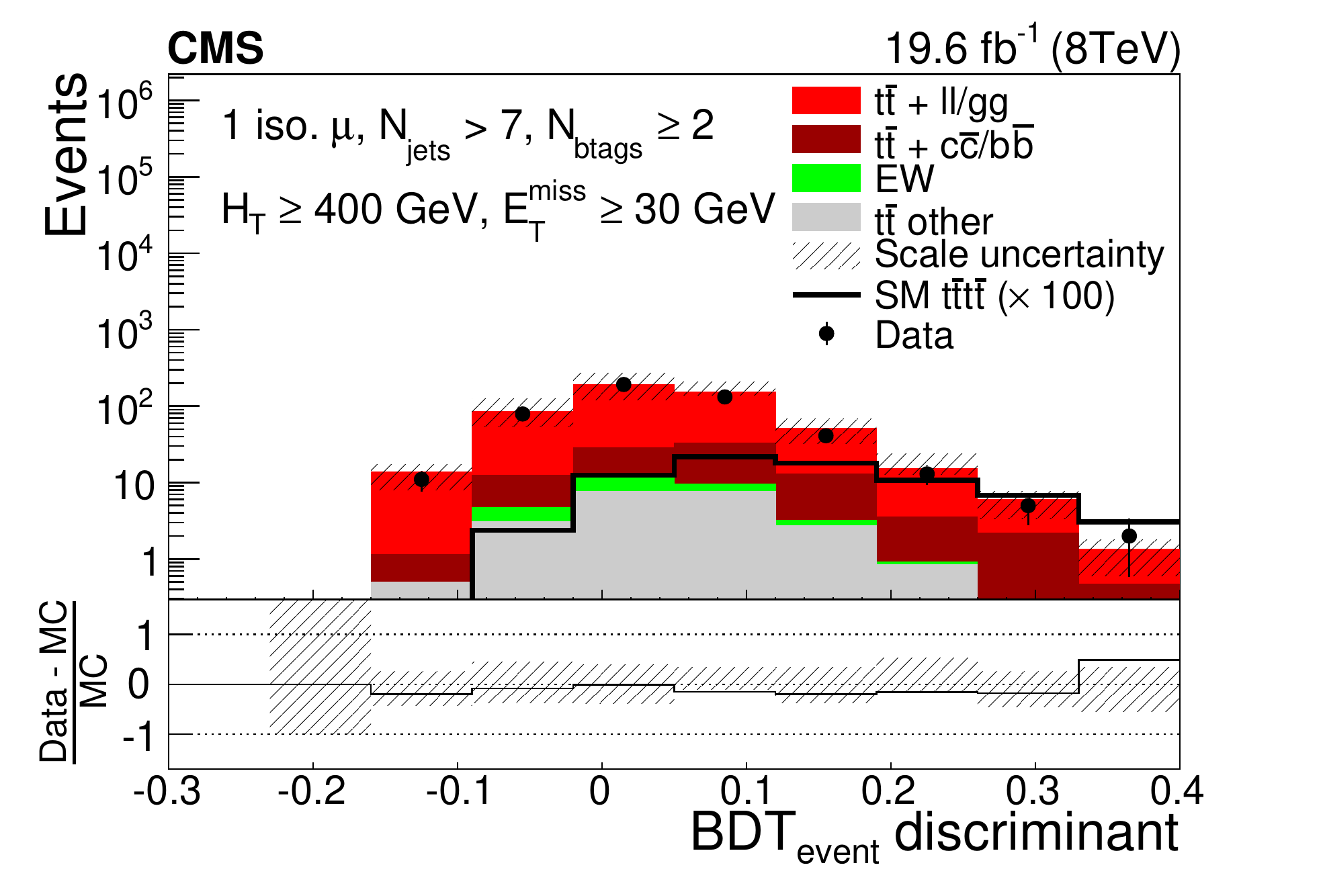}}
\subfigure[]{\includegraphics[width=.49\textwidth]{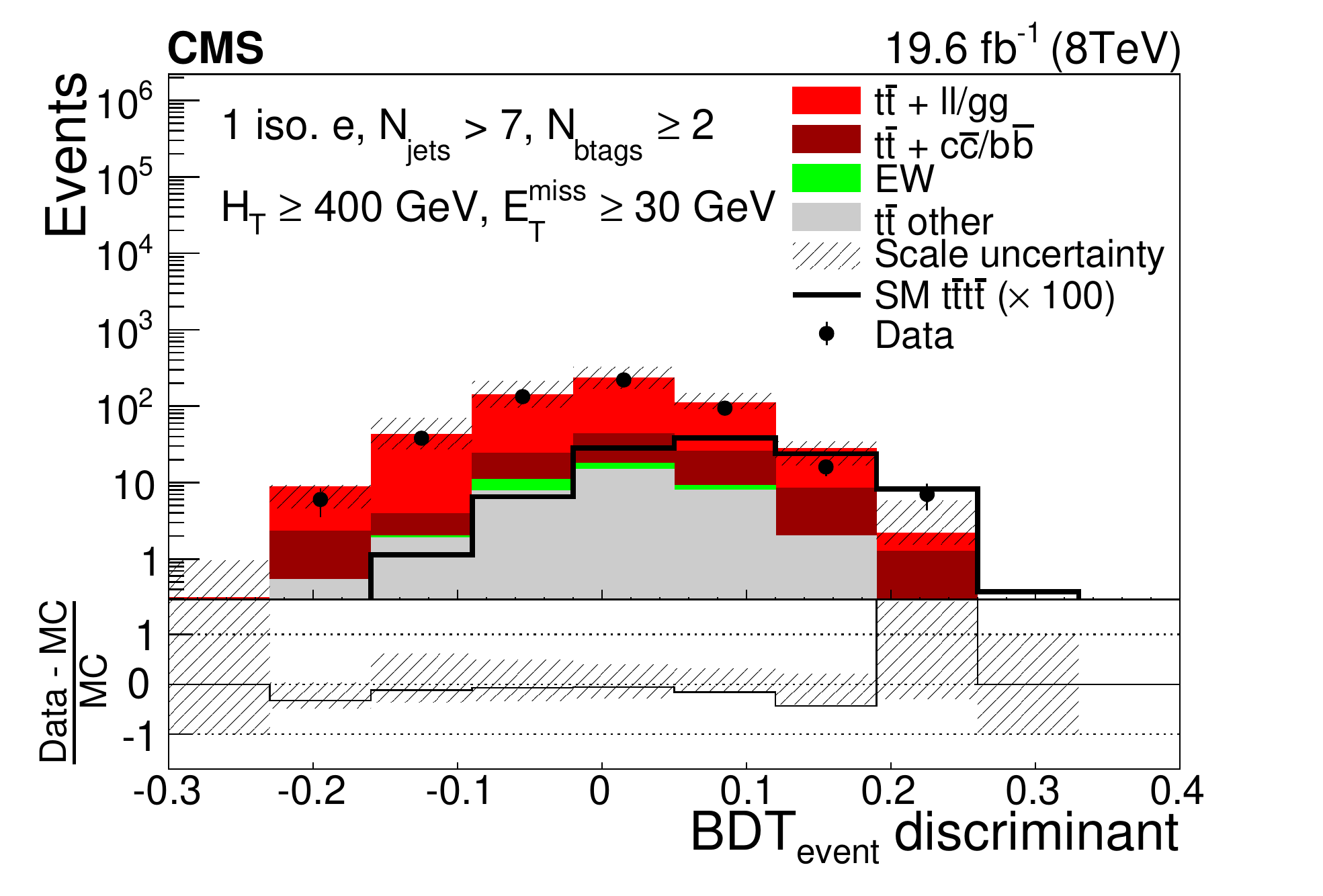}} \\
\caption{The distribution in the BDT\subb{event} discriminant for data and simulation in events with $N_\text{jets} = 6$ for the $\mu$ + jets and e + jets channels in (a) and (b), respectively, and the same in events with $N_\text{jets} = 7$ in (c) and (d), and in events with $N_\text{jets} > 7$ in (e) and (f). The ratios plotted at the bottom of each panel reflect the percent differences between data and MC events. The hatched areas reflect the changes in the calculated predictions produced by factors of two and one half changes in the factorisation and renormalisation scales (see Section~\ref{sec:limitssys}).}
\label{fig:BC1}
\end{center}
\end{figure}

The ten variables described in Sections \ref{sec:topcon}, \ref{sec:jetact}, and \ref{sec:bcon} are combined using a second, event-level  BDT (BDT\subb{event}). To maximise sensitivity, the events are divided into three categories corresponding to $N_\text{jets}$ = 6, 7 and, $>$ 7, where the $N_\text{jets}$ = 6 category is used as a sideband region to constrain the \ttbar background in the calculation of limits on \tttt production. In Fig. \ref{fig:BC1}, distributions in the BDT\subb{event} discriminant are shown in data and in MC for each of these categories.

\section{Systematic uncertainties and limits on \texorpdfstring{\tttt}{t t-bar t t-bar} production \label{sec:limitssys}}
The systematic uncertainties considered in this analysis are separated into two categories: (i) those that affect the normalisations of the BDT\subb{event} discriminant distributions of both signal and backgrounds and (ii) those that affect the form of the distributions of just the backgrounds. The normalisations are affected by the uncertainty in integrated luminosity of the data and the theoretical cross sections of the signal and background processes. An uncertainty in the integrated luminosity of 2.6\% is included \cite{CMS-PAS-LUM-13-001}. The uncertainty in the \ttbar cross section is expected to dominate, and is taken from Ref. \cite{PhysRevLett.110.252004} as $^{+ 2.5\%}_{ - 3.4\%}$~(factorisation and renormalisation scales) and $^{+ 2.5\%}_{ - 2.6\%}$~(PDF). A range of theoretical and experimental sources of systematic uncertainty can affect the form of the distributions of the BDT\subb{event} discriminant. As \ttbar is the dominant background, systematic effects on the form of the distributions are considered only for \ttbar events. The impact of contributions from higher-order corrections in the \ttbar simulation is quantified by comparing alternative \ttbar samples that are generated with the renormalisation and factorisation scales simultaneously changed up and down by a factor of two relative to the nominal \ttbar sample. The matching of partons originating from the matrix element to the jets from the parton showers is performed according to the MLM prescription \cite{Mangano:2006rw}. The uncertainty arising from this prescription is estimated by changing the minimum \kt measure between partons by factors of 0.5 and 2.0 and the jet matching threshold by factors of 0.75 and 1.5. To evaluate the uncertainty due to imperfect knowledge of the JES, JER, b tagging, and lepton-identification efficiencies, and the cross section for minimum-bias production used in the pileup-reweighting procedure in simulation, the input value of each parameter is changed by $\pm$1 standard deviation of its uncertainty. A systematic uncertainty due to the imperfect knowledge of the contribution from the $\ttbar\bbbar$ component in \ttbar events is also estimated.  As mentioned previously, a correction is applied to the \ttbar simulation to reproduce the observed \cite{CMS-PAS-TOP-13-010} ratio of $\sigma_{\ttbar\bbbar}$ to $\sigma_{\ttbar jj}$. The systematic uncertainty associated with the imperfect knowledge of this ratio is estimated by changing this correction by $\pm$50\%.

No significant excess of events to represent SM \tttt production is observed above the background prediction. Therefore, an upper limit on $\sigma_{\tttt}$ is set by performing a simultaneous maximum likelihood fit to the distributions in the BDT\subb{event} discriminant for signal and background in the six event categories described in Section \ref{sec:bdt}. The systematic uncertainties in the normalisation and the form of the distributions of the discriminant are accommodated by incorporating the uninteresting nuisance parameters into the fit. The contribution of the nuisance parameters to the likelihood function are modelled using log-normal functions for normalisations and Gaussian functions for forms of the distributions. The functions have widths that correspond to the $\pm$1 standard deviation changes of the systematic sources described in the previous section. Statistical uncertainties in the simulation are taken into account by applying a ``lightweight" version of the Beeston and Barlow method \cite{BeestonBarlow} where one nuisance parameter is associated with the estimate of the total simulation and the  statistical uncertainty in each bin. The best-fit values of the nuisance parameters show only statistically insignificant deviations from their input values. In particular, the best-fit value of the parameter corresponding to the $\ttbar$$\bbbar$ correction is consistent with the result obtained in Ref. \cite{CMS-PAS-TOP-13-010}.

The modified frequentist CL\subb{s} approach \cite{Junk:1999kv,Read:2002hq} using the asymptotic approximation is adopted to measure the upper limit using the \textsc{RooStats} package \cite{Cowan:2010js,RooStats}. The limit calculated at a 95\% confidence level (CL) on the production cross section $\sigma_{\tttt}$ is 32\unit{fb}, where a limit of $32\pm{17}$\unit{fb} is expected. These limits are approximately $25 \times\sigma^{\text{SM}}_{\tttt}$.

\section{Summary}
A search for events containing four top quarks was performed using data collected with the CMS detector in lepton + jets final states at $\sqrt{s} = 8$\TeV, corresponding to an integrated luminosity of \ourLumi. The analysis had three stages. First, a baseline selection was used to select signal events while suppressing backgrounds. Second, to further discriminate between signal and background, an event classification scheme based on a BDT algorithm was defined to exploit differences in the multiplicity of top quarks, jet activity, and the multiplicity of bottom quarks. Third, a simultaneous maximum likelihood fit of the BDT\subb{event} discriminant distributions was performed, from which an upper limit on $\sigma_{\tttt}$ of 32\unit{fb} was calculated at a 95\% CL, where a limit of $32\pm{17}$\unit{fb} was expected. These limits are approximately $25 \times \sigma^{\text{SM}}_{\tttt}$. This result raises the prospect of the direct observation of SM \tttt in future CMS data at the higher centre-of-mass energies of 13 and 14\TeV, where $\sigma^{\text{SM}}_{\tttt}$ is predicted to be $\approx$9\unit{fb} and 15\unit{fb}, respectively \cite{threeTop,Alwall:2011uj}. Furthermore, this result has the potential to constrain BSM theories producing \tttt final states with kinematics similar to the SM process, and enhance the discovery reach of BSM searches where SM production of \tttt constitutes a possible background.

\begin{acknowledgments}
We congratulate our colleagues in the CERN accelerator departments for the excellent performance of the LHC and thank the technical and administrative staffs at CERN and at other CMS institutes for their contributions to the success of the CMS effort. In addition, we gratefully acknowledge the computing centres and personnel of the Worldwide LHC Computing Grid for delivering so effectively the computing infrastructure essential to our analyses. Finally, we acknowledge the enduring support for the construction and operation of the LHC and the CMS detector provided by the following funding agencies: BMWFW and FWF (Austria); FNRS and FWO (Belgium); CNPq, CAPES, FAPERJ, and FAPESP (Brazil); MES (Bulgaria); CERN; CAS, MoST, and NSFC (China); COLCIENCIAS (Colombia); MSES and CSF (Croatia); RPF (Cyprus); MoER, ERC IUT and ERDF (Estonia); Academy of Finland, MEC, and HIP (Finland); CEA and CNRS/IN2P3 (France); BMBF, DFG, and HGF (Germany); GSRT (Greece); OTKA and NIH (Hungary); DAE and DST (India); IPM (Iran); SFI (Ireland); INFN (Italy); NRF and WCU (Republic of Korea); LAS (Lithuania); MOE and UM (Malaysia); CINVESTAV, CONACYT, SEP, and UASLP-FAI (Mexico); MBIE (New Zealand); PAEC (Pakistan); MSHE and NSC (Poland); FCT (Portugal); JINR (Dubna); MON, RosAtom, RAS and RFBR (Russia); MESTD (Serbia); SEIDI and CPAN (Spain); Swiss Funding Agencies (Switzerland); MST (Taipei); ThEPCenter, IPST, STAR and NSTDA (Thailand); TUBITAK and TAEK (Turkey); NASU and SFFR (Ukraine); STFC (United Kingdom); DOE and NSF (USA).

Individuals have received support from the Marie-Curie programme and the European Research Council and EPLANET (European Union); the Leventis Foundation; the A. P. Sloan Foundation; the Alexander von Humboldt Foundation; the Belgian Federal Science Policy Office; the Fonds pour la Formation \`a la Recherche dans l'Industrie et dans l'Agriculture (FRIA-Belgium); the Agentschap voor Innovatie door Wetenschap en Technologie (IWT-Belgium); the Ministry of Education, Youth and Sports (MEYS) of the Czech Republic; the Council of Science and Industrial Research, India; the HOMING PLUS programme of Foundation for Polish Science, cofinanced from European Union, Regional Development Fund; the Compagnia di San Paolo (Torino); the Consorzio per la Fisica (Trieste); MIUR project 20108T4XTM (Italy); the Thalis and Aristeia programmes cofinanced by EU-ESF and the Greek NSRF; and the National Priorities Research Program by Qatar National Research Fund. 
\end{acknowledgments}

\bibliography{auto_generated}   

\cleardoublepage \appendix\section{The CMS Collaboration \label{app:collab}}\begin{sloppypar}\hyphenpenalty=5000\widowpenalty=500\clubpenalty=5000\input{TOP-13-012-authorlist.tex}\end{sloppypar}
\end{document}

%% file: TOP-13-012-authorlist.tex
\textbf{Yerevan Physics Institute,  Yerevan,  Armenia}\\*[0pt]
V.~Khachatryan, A.M.~Sirunyan, A.~Tumasyan
\vskip\cmsinstskip
\textbf{Institut f\"{u}r Hochenergiephysik der OeAW,  Wien,  Austria}\\*[0pt]
W.~Adam, T.~Bergauer, M.~Dragicevic, J.~Er\"{o}, C.~Fabjan\cmsAuthorMark{1}, M.~Friedl, R.~Fr\"{u}hwirth\cmsAuthorMark{1}, V.M.~Ghete, C.~Hartl, N.~H\"{o}rmann, J.~Hrubec, M.~Jeitler\cmsAuthorMark{1}, W.~Kiesenhofer, V.~Kn\"{u}nz, M.~Krammer\cmsAuthorMark{1}, I.~Kr\"{a}tschmer, D.~Liko, I.~Mikulec, D.~Rabady\cmsAuthorMark{2}, B.~Rahbaran, H.~Rohringer, R.~Sch\"{o}fbeck, J.~Strauss, A.~Taurok, W.~Treberer-Treberspurg, W.~Waltenberger, C.-E.~Wulz\cmsAuthorMark{1}
% ...
\vskip\cmsinstskip
% ...
1:~~Also at Vienna University of Technology, Vienna, Austria\\
2:~~Also at CERN, European Organization for Nuclear Research, Geneva, Switzerland\\
% ...
